\documentclass[twoside]{elsart}

\usepackage{amsmath,epsfig,amsfonts,amssymb,latexsym}

\begin{document}
\date{\today}
\begin{frontmatter}
\title			{Mixtures of compound Poisson processes as models
of tick-by-tick financial data}
 
\author{Enrico Scalas}

\address{Dipartimento di Scienze e Tecnologie Avanzate, 
			Universit\`a del Piemonte Orientale, 
			via Bellini 25 g, 
			I--15100 Alessandria, Italy}

\begin{abstract}
A model for the phenomenological description of tick-by-tick share prices in a stock
exchange is introduced. It is based on mixtures of compound Poisson processes. Preliminary results based on Monte Carlo simulation show that this model can reproduce various stylized facts.
\end{abstract}

\begin{keyword}
Waiting-time; Duration; random walk; statistical finance;
\\ {\it Corresponding author}: Enrico Scalas ({\tt scalas@unipmn.it}), 
\\ url: {\tt www.econophysics.org}
\end{keyword}

\end{frontmatter}

\def\eg{{\it e.g.}\ } \def\ie{{\it i.e.}\ }
\def\sg{\hbox{sign}\,}
\def\sgn{\hbox{sign}\,}
\def\sign{\hbox{sign}\,}
\def\e{\hbox{e}}
\def\exp{\hbox{exp}}
\def\ds{\displaystyle}
\def\dis{\displaystyle}
\def\q{\quad}    \def\qq{\qquad}
\def\lan{\langle}\def\ran{\rangle}
\def\l{\left} \def\r{\right}
\def\lra{\Longleftrightarrow}
\def\arg{\hbox{\rm arg}}
\def\d{\partial}
 \def\dr{\partial r}  \def\dt{\partial t}
\def\dx{\partial x}   \def\dy{\partial y}  \def\dz{\partial z}
\def\rec#1{{1\over{#1}}}
\def\log{\hbox{\rm log}\,}
\def\erf{\hbox{\rm erf}\,}     \def\erfc{\hbox{\rm erfc}\,}
\def\F{\hbox{F}\,}
\def\NN{\hbox{\bf N}}
\def\RR{\hbox{\bf R}}
\def\CC{\hbox{\bf C}}
\def\ZZ{\hbox{\bf Z}}
\def\II{\hbox{\bf I}}


\section{Introduction}

Continuous time random walks (CTRWs) were introduced in Physics by Montroll and Weiss as a model for 
single-particle (tracer) diffusion \cite{montroll65}. An instance of CTRW, the normal compound Poisson 
process, had already been used in the probabilistic theory of insurance ruin since the beginning of the 
XXth Century \cite{lundberg03,cramer30}.

The seminal paper of Montroll and Weiss has been followed by many studies focusing on anomalous 
relaxation and anomalous diffusion. This is the subject of two recent reviews by Metzler and Klafter 
\cite{metzler00,metzler04}.

The present author has recently reviewed the applications of CTRWs to Finance and Economics \cite{scalas06a}. 
These applications were triggered by a series of papers on finance and fractional calculus 
\cite{scalas00,mainardi00,gorenflo01}, but the reader is referred to
ref. \cite{press67} for an early application of the normal compound Poisson process to financial data.

The recent research of the present author has focused on the behaviour of waiting times (also known as 
durations) between trades and order in financial markets \cite{mainardi00,raberto02,scalas04,scalas06b}. 
It turned out that interorder and intertrade waiting times are not exponentially distributed. Therefore, 
the jump process of tick-by-tick prices is non-Markovian \cite{mainardi00}.

In an article within this issue \cite{bianco06}, Bianco and Grigolini apply a new method
to verify whether the intertrade waiting time process is a genuine renewal process 
\cite{cox67,allegrini06a,allegrini06b}. This was assumed by the CTRW hypothesis in \cite{scalas00}. 
They find that intertrade waiting times follow a renewal process.

Here, inspired by the work of Edelman and Gillespie \cite{gillespie99,edelman00}, a phenomenological model for 
intraday tick-by-tick financial data is presented. It is able to reproduce some important stylized facts. The 
paper is organized as follows. Section 2 contains an outline of the theory of CTRWs. Section 3 contains a 
description of the model as well as a discussion on results from Monte Carlo simulations.

\section{Outline of theory}

\subsection{Basic definitions}

CTRWs are point processes with reward. The point process is characterized by  a sequence of independent 
identically distributed (i.i.d.) positive random variables $\tau_i$, which can be interpreted as waiting 
times between two consecutive events:
\begin{equation}
\label{timewalk}
t_n = t_0 + \sum_{i=1}^{n} \tau_i; \; \; t_n - t_{n-1} = \tau_n; \; \; n=1, 2, 3, \ldots;
\; \; t_0 = 0. \end{equation} 
The rewards are i.i.d. not necessarily positive random variables: $\xi_i$. In the usual physical intepretation, 
they represent the jumps of a random walker, and they can be $n$-dimensional vectors. In this paper, only the 
1-dimensional case is studied for a continuous random variable, but the extension of many results to the 
$n$-dimensional case and to a lattice is straightforward. The position $x$ of the walker at time $t$ is 
(with $N(t) = \max \{ n:\;t_{n} \leq t \}$ and $x(0)=0$):
\begin{equation}
\label{jumpwalk}
x(t) = \sum_{i=1}^{N(t)} \xi_i.
\end{equation}
CTRWs are rather good and general phenomenological models for diffusion, including anomalous diffusion, provided 
that the time of residence of the walker is much greater than the time it takes to make a jump. In fact, in this 
formalism, jumps are instantaneous.

The financial interpretation of the random variables is straightforward. If trades take place in a continuous 
double auction, both price variations and waiting times (also called durations) between two consecutive trades 
are random variables. If $S(t)$ is the price of an asset at time $t$ defined according to the previous tick 
interpolation procedure, $S(t) = S(t_i)$ where $t_i$ is the time instant at which the last trade took place, 
then the price process can be considered as a pure jump stochastic process in continuous time. In finance, it 
is better to work with returns rather that prices. If $S(0)$ is the price at time $t=0$, then the variable 
$x(t) = \log (S(t)/S(0))$ is called the {\em log-return} or, better, the {\em log-price}. This variable 
is analogous to the position of the walker in the physical interpretation. In the financial intepretation 
the jump random variables $\xi_i$ are tick-by-tick log returns and they coincide with the difference between 
two consecutive log prices, whereas the waiting times or durations $\tau_i$ denote the elapsed time between 
two consecutive trades. 

In general, jumps and waiting times are not independent from each other. In any case, a CTRW is characterized 
by the joint probability density $\varphi (\xi, \tau)$ of jumps and waiting times; 
$\varphi (\xi, \tau) \, d \xi \, d\tau$ is the probability of a jump to be in the interval 
$(\xi, \xi+ \, d\xi)$ and of a waiting time to be in the interval $(\tau, \tau+ \, d\tau)$. The following 
integral equation gives the probability density, $p(x,t)$, for the walker being in position $x$ at time $t$, 
conditioned on the fact that it was in position $x=0$ at time $t=0$:
\begin{equation}
\label{masterequation}
p(x,t) =  \delta (x)\, \Psi(t) +
   \int_0^t \, 
 \int_{-\infty}^{+\infty}  \varphi(x-x',t-t')\, p(x',t')\, dt'\,dx',
\end{equation}
where $\Psi(\tau)$ is the so-called survival function. $\Psi(\tau)$ is related to the marginal waiting-time 
probability density $\psi(\tau)$. The two marginal densities $\psi(\tau)$ and $\lambda(\xi)$ are:
\begin{eqnarray}
\label{marginal}
\psi(\tau) & = & \int_{-\infty}^{+\infty} \varphi(\xi, \tau) \, d \xi \nonumber \\
\lambda(\xi) & = & \int_{0}^{\infty} \varphi(\xi, \tau) \, d \tau,
\end{eqnarray}
and the survival function $\Psi(\tau)$ is defined as:
\begin{equation}
\label{survival}
\Psi(\tau) = 1 - \int_{0}^{\tau} \psi (\tau') \, d \tau' = \int_{\tau}^{\infty} \psi (\tau') \, d \tau'.
\end{equation}

The integral equation, eq. (\ref{masterequation}) is linear and it can be solved in the Laplace-Fourier domain. 
The Laplace transform, $\widetilde{g}(s)$ of a (generalized) function $g(t)$ is defined as:
\begin{equation}
\label{laplacetransform}
\widetilde{g}(s) = 
\int_{0}^{+ \infty} dt \; 
\hbox{e}^{ -st} \, g(t)\,,
\end{equation}
whereas the Fourier transform of a (generalized) function $f(x)$ is defined as:
\begin{equation}
\label{fouriertransform}
\widehat {f}(\kappa) = 
\int_{- \infty}^{+ \infty} dx \, 
\hbox{e}^{i \kappa x} \, f(x)\,.
\end{equation}
A generalized function is a distribution (like Dirac's $\delta$) in the sense of S. L. Sobolev and 
L. Schwartz \cite{gelfand58}.

One gets:
\begin{equation}
\label{gensol}
\widetilde{\widehat p}(\kappa, s) = \widetilde \Psi(s)\, \frac{1}{1-\widetilde{\widehat \varphi}(\kappa, s)},
\end{equation}
or, in terms of the density $\psi(\tau)$:
\begin{equation}
\label{vargensol}
\widetilde{\widehat p}(\kappa, s) = \frac{1 -\widetilde \psi(s)}{s}\, \frac{1}{1-\widetilde{\widehat \varphi}(\kappa, s)},
\end{equation}
as, from eq. (\ref{survival}), one has:
\begin{equation}
\label{survivallt}
\Psi(s) = \frac{1 -\widetilde \psi(s)}{s}.
\end{equation} 
In order to obtain $p(x,t)$, it is then necessary to invert its Laplace-Fourier transform 
$\widetilde{\widehat p}(\kappa, s)$. Analytic solutions are quite important, as they provide a 
benchmark for testing numerical inversion methods. In the next section, an explicit analytic solution 
for a class of continuous-time random walks with anomalous relaxation behaviour will be presented. 
It will be necessary to restrict oneself to the uncoupled case, in which jumps and waiting-times 
are not correlated. 

\subsection{The normal compound Poisson process}

In this section, the solution of eq. (\ref{masterequation}) will be derived in the uncoupled case 
where the joint probability density of jumps and durations can be factorized in term of its marginals. 
After the derivation of a genearal formula for $p(x,t)$, this will be specialized to the case of the 
normal compound Poisson process (NCPP).

If jump sizes do not depend on waiting times, the joint probability density for jumps and waiting times 
can be written as follows:
\begin{equation}
\label{factorization}
\varphi(\xi, \tau) = \lambda(\xi) \psi (\tau)
\end{equation}
with the normalization conditions $\int d \xi \lambda (\xi) = 1$ and $\int d \tau \psi(\tau)=1$.

In this case the integral master equation for $p(x,t)$ becomes:
\begin{equation}
\label{uncoupledreal}
p(x,t) =  \delta (x)\, \Psi(t) +
   \int_0^t   \psi(t-t') \, \left[
 \int_{-\infty}^{+\infty}  \lambda(x-x')\, p(x',t')\, dx'\right]\,dt'
\end{equation}
This equation has a well known general explicit solution in terms of $P(n,t)$, the probability of $n$ 
jumps occurring up to time $t$, and of the $n$-fold convolution of the jump density, $\lambda_n (x)$: 
\begin{equation}
\label{jumpconv}
\lambda_n (x) =
\int_{-\infty}^{+\infty} \, \int_{-\infty}^{+\infty} \ldots \int_{-\infty}^{+\infty} \,d \xi_{n-1} 
d \xi_{n-2} \ldots d \xi_1 \lambda(x - \xi_{n-1}) \lambda (\xi_{n-1}-\xi_{n-2}) \ldots \lambda(\xi_1).
\end{equation}
Indeed, $P(n,t)$ is given by:
\begin{equation}
\label{Poisson1}
P(n,t) = \int_{0}^{t} \psi_n (t - \tau) \Psi(\tau) \, d \tau
\end{equation}
where $\psi_n (\tau)$ is the $n$-fold convolution of the waiting-time density: 
\begin{equation}
\label{timeconv}
\psi_n (\tau) = \int_{0}^{\tau} \int_{0}^{\tau_{n-1}} \ldots \int_{0}^{\tau_1}
\, d \tau_{n-1} d \tau_{n-2} \ldots d \tau_1 \psi(t-\tau_{n-1}) \psi(\tau_{n-1} - \tau_{n-2}) \ldots \psi(\tau_1).   
\end{equation}
The $n$-fold convolutions defined above are probability density functions for the sum of $n$ variables.

The Laplace transform of $P(n,t)$, $\widetilde P(n,s)$, reads:
\begin{equation}
\label{Poisson1lt}
\widetilde P(n,s) = [ \widetilde \psi (s) ]^n \widetilde \Psi(s)
\end{equation}
By taking the Fourier-Laplace transform of eq. (\ref{uncoupledreal}), one gets:
\begin{equation}
\label{montroll}
\widetilde{\widehat p}(\kappa, s) = \widetilde \Psi(s)\,
\frac{1}{1- \widetilde \psi(s) \widehat \lambda(\kappa)}\,.
\end{equation}
But, recalling that $|\lambda(\kappa)| < 1$ and $|\psi(s)| < 1$, if $\kappa \not= 0$ and 
$s \not= 0$, eq. (\ref{montroll}) becomes:
\begin{equation}
\label{series1}
\widetilde{\widehat p}(\kappa, s) = \widetilde \Psi(s)\, \sum_{n=0}^{\infty} 
[\widetilde \psi(s) \widehat \lambda(\kappa)]^n \,;
\end{equation}
this gives, inverting the Fourier and the Laplace transforms and taking into account eqs. (\ref{jumpconv}) 
and (\ref{Poisson1}):
\begin{equation}
\label{series2}
p(x,t) = \sum_{n=0}^{\infty} P(n,t) \lambda_n (x)
\end{equation}
Eq. (\ref{series2}) can also be used as the starting point to derive eq. (\ref{uncoupledreal}) via the 
transforms of Fourier and Laplace, as it describes a jump process subordinated to a renewal process.

A remarkable analytic solution is available when the waiting-time probability density function has the 
following exponential form:
\begin{equation}
\label{exponential1}
\psi(\tau) = \mu \hbox{e}^{-\mu \tau}.
\end{equation}
Then, the survival probability is $\Psi (\tau) = \hbox{e}^{-\mu \tau}$ and the probability of $n$ jumps 
occurring up to time $t$ is given by the Poisson distribution:
\begin{equation}
\label{Poissondistribution}
P(n,t) = \frac{(\mu t)^n}{n!} \, \hbox{e}^{-\mu t}.
\end{equation}
This is the only Markovian case, and equation (\ref{series2}) becomes:
\begin{equation}
\label{Lundbergsolution}
p(x,t) = \sum_{n=0}^{\infty} \frac{(\mu t)^n}{n!} \, \hbox{e}^{-\mu t} \lambda_{n}(x).
\end{equation}
If $\lambda(\xi)$ follows the normal distribution $N(\xi;\bar{\xi},\sigma_{\xi})$, then the n-fold convolution 
is given by: $\lambda_{n}(x) = N(x; n \bar{\xi}, \sqrt{n} \sigma_{\xi})$.

Given a series of empirical tick-by-tick log returns, $\{ \xi_i \}_{i}^{M}$, as well as durations, 
$\{ \tau_i \}_{i}^{M}$, one can directly evaluate the three parameters 
$\mu$: the activity of the Poisson process, $\bar{\xi}$: the average of log-returns, and
$\sigma_{\xi}$: the standard deviation of log-returns by means of suitable estimators \cite{press67}.

However, the normal compound Poisson process is not able to reproduce the following stylized facts 
on high frequency data:

\begin{enumerate}

\item

The empirical distribution of log-returns is leptokurtic, whereas the NCPP assumes a mesokurtic 
(actually normal) distribution.

\item

The empirical distribution of durations is non-exponential with excess standard deviation 
\cite{engle97,engle98,mainardi00,raberto02,scalas04}, whereas the NCPP assumes an exponential distribution.

\item

The autocorrelation of absolute log-returns decays slowly \cite{raberto02}, whereas the NCPP assumes i.i.d. log-returns.

\item

Log-returns and waiting times are not independent \cite{raberto02,meerschaert06}, 
whereas the NCPP assumes their independence.

\item

Volatility and activity vary during the trading day \cite{bertram04}, whereas the NCPP assumes they are constant.

\end{enumerate}

\section{Mixtures of normal compound Poisson processes}

\subsection{Definition}

It is possible to overcome the above shortcomings by using a suitable mixture of NCPPs. During a 
trading day, the volatility and the activity are higher at the opening of the market, then they decrease 
at midday and they increase again at market closure \cite{bertram04}. If the trading day can be divided 
into $T$ intervals of constant activity $\{ \mu_i \}_{i=1}^{T}$, then the waiting-time distribution is a 
mixture of exponential distributions and its probability density can be written as:
\begin{equation}
\label{poissonmixture}
\psi (\tau) = \sum_{i=1}^{T} a_i \mu_i \hbox{e}^{-\mu_i \tau},
\end{equation}
where $\{ a_i \}_{i=1}^{T}$ is a set of suitable weights. The activity seasonality can be mimicked by 
values of $\mu_i$ that decrease towards midday and then increase again towards market closure. In order 
to reproduce the correlation between volatility and activity, one can assume that:
\begin{equation}
\label{corrvolact}
\sigma_{\xi,i} = c \mu_i
\end{equation}
where $c$ is a suitable constant. Future work will be devoted to an analytical study of this model as well as to 
further empirical investigations on model validation. Below, the results of a simulation performed with the 
model are presented, and the performance of the model with respect to the stylized facts is discussed.

\subsection{Results}

A Monte Carlo simulation of the model described in the previous subsection has been performed by considering 
a trading day divided into ten intervals of constant activity with 
$\{\mu_i \}_{i=1}^{10}$ = $1/10,1/20,1/30,1/40,1/50,1/40,1/30,1/20$, $1/15,1/10$ s$^{-1}$. 
For each value of $\mu_i$, 100 exponentially distributed waiting times have extracted as well as 
100 normally distributed log-returns with zero average and 
$\sigma_{\xi,i} = 0.001 \cdot \mu_i$. Therefore, there are 1000 values of waiting times and log returns 
in a trading day, representing a rather liquid share. The opening price is set to 100 arbitrary units (a.u.). 
In Fig. 1, a sample path is plotted for the price as a function of trading time. Fig. 2 and Fig. 3 represent 
the tick-by-tick time series of log-returns and waiting times respectively. For this particular simulation, 
the effect of variable activity and volatility can be detected by direct eye inspection.

\begin{figure}
\begin{center}
\mbox{\epsfig{file=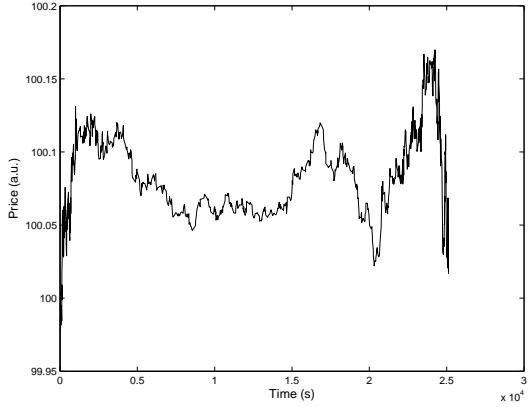,width=7.cm,angle=0}}
\end{center}
\caption{Simulated price as a function of transaction time. The initial price is set to 100 arbitrary 
units (a.u.). Simulation times are measured in seconds}
\label{f.1}
\end{figure}

\begin{figure}
\begin{center}
\mbox{\epsfig{file=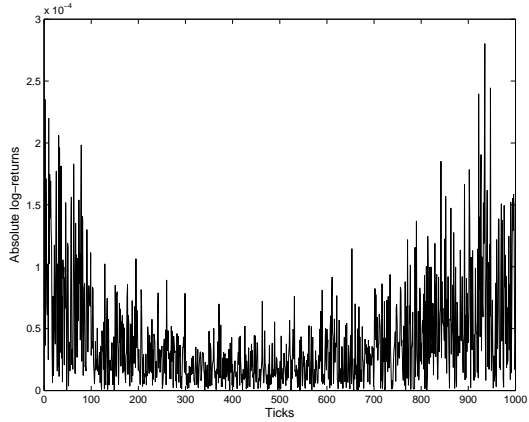,width=7.cm,angle=0}}
\end{center}
\caption{A simulated absolute log-return series.}
\label{f.2}
\end{figure}

\begin{figure}
\begin{center}
\mbox{\epsfig{file=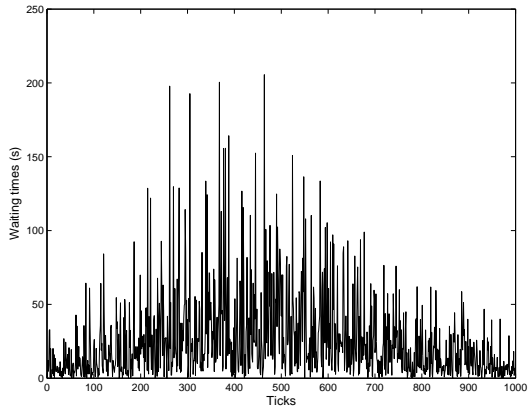,width=7.cm,angle=0}}
\end{center}
\caption{A simulated waiting time series.}
\label{f.3}
\end{figure}

In order to show that this model is able to reproduce the stylized facts described above, another set of 
figures is presented in the following. In Fig. 4 the empirical complementary cumulative distribution function 
is plotted for absolute tick-by-tick log
returns. For comparison, the Gaussian fit with the same standard deviation of the 1000 log-returns is given 
by a solid line. This distribution has fat tails, is leptokurtic and the kurtosis is equal to 6.

\begin{figure}
\begin{center}
\mbox{\epsfig{file=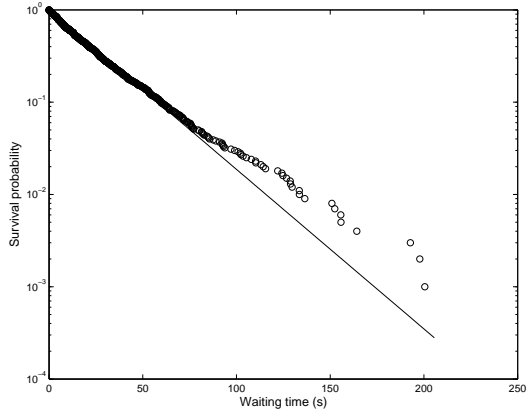,width=7.cm,angle=0}}
\end{center}
\caption{Empirical complementary cumulative distribution function for absolute log returns (circles). 
The solid line is a Gaussian fit.}
\label{f.4}
\end{figure}

The empirical complementary cumulative distribution function for intertrade durations is given in Fig. 5. 
The solid line is the single exponential fit to the simulated data. There is excess standard deviation: 
the standard deviation of waiting times is 29 s, whereas the average waiting time is 25 s.

\begin{figure}
\begin{center}
\mbox{\epsfig{file=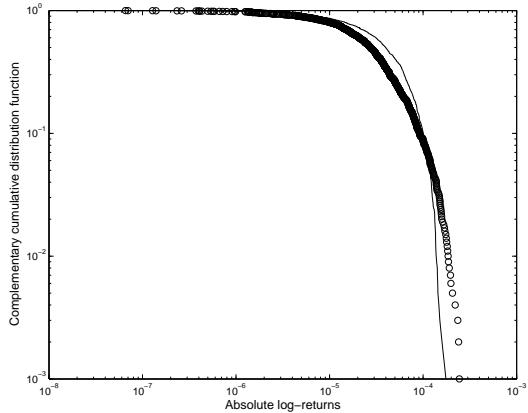,width=7.cm,angle=0}}
\end{center}
\caption{Empirical survival probability.}
\label{f.5}
\end{figure}

Fig. 6 shows the slow decay of the autocorrelation of absolute log-returns related to volatility 
clustering, whereas signed log-returns are zero already at the second lag.

\begin{figure}
\begin{center}
\mbox{\epsfig{file=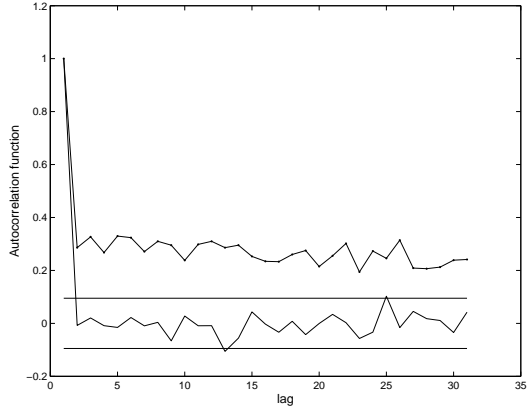,width=7.cm,angle=0}}
\end{center}
\caption{Estimate of the autocorrelation function for absolute log-returns (.-),
and signed log-returns (-). The solid horizontal lines represent the statistical
zero level ( $\pm$ 3/$\sqrt{1000}$).}
\label{f.6}
\end{figure}

In conclusion, the model based on mixtures of normal compound Poisson processes incorporates 
variable daily activity, as well as the dependence between durations and tick-by-tick log-returns 
via eq. (\ref{corrvolact}). It is then able to replicate the following stylized facts:

\begin{itemize}

\item

The empirical distribution of log-returns is leptokurtic;

\item

the empirical distribution of durations is non-exponential with excess standard deviation;

\item

The autocorrelation of absolute log-returns decays slowly.

\end{itemize}

Work is currently in progress to empirically validate the model \cite{scalas06c}.

\section*{ACKNOWLEDGEMENTS}

The authors acknowledges an interesting discussion with Peter Buchen and Tom Gillespie. 
He is indebted to Rudolf Gorenflo and Francesco Mainardi with whom he developed the application 
of continuous time random walks to finance. This work has been supported by the Italian MIUR 
grant "Dinamica di altissima frequenza nei mercati finanziari".

\end{document}